\documentclass[%
 reprint,
]{revtex4-1}

\usepackage{graphicx}
\usepackage{dcolumn}
\usepackage{bm}
\usepackage{amsmath}

\begin{document}

\preprint{APS/123-QED}

\title{Improving local clustering based top-L link prediction methods \\via asymmetric link clustering information}

\author{Zhihao Wu }
\email{zhwu@bjtu.edu.cn}
 
\author{Youfang Lin}
\email{yflin@bjtu.edu.cn}

\author{Yiji Zhao}

\affiliation{%
Beijing Key Lab of Traffic Data Analysis and Mining,\\
School of Computer and Information Technology,\\
Beijing Jiaotong Univerisy, Beijing 100044, People's Republic of China
}%

	\author{Hongyan Yan}
	
	\affiliation{User and Market Research Department,\\
	China Mobile Research Institute, Beijing 100032
	}


\begin{abstract}
Networks can represent a wide range of complex systems, such as social, biological and technological systems. Link prediction is one of the most important problems in network analysis, and has attracted much research interest recently. Many link prediction methods have been proposed to solve this problem with various technics. We can note that clustering information plays an important role in solving the link prediction problem. In previous literatures, we find node clustering coefficient appears frequently in many link prediction methods. However, node clustering coefficient is limited to describe the role of a common-neighbor in different local networks, because it can not distinguish different clustering abilities of a node to different node pairs. In this paper, we shift our focus from nodes to links, and propose the concept of asymmetric link clustering (ALC) coefficient. Further, we improve three node clustering based link prediction methods via the concept of ALC. The experimental results demonstrate that ALC-based methods outperform node clustering based methods, especially achieving remarkable improvements on food web, hamster friendship and Internet networks. Besides, comparing with other methods, the performance of ALC-based methods are very stable in both globalized and personalized top-L link prediction tasks. \\
\\
\end{abstract}

\maketitle


\section{INTRODUCTION}

Networks have been widely used to represent a large variety of complex systems, such as social, biological and technological systems \cite{albert2002statistical,boccaletti2006complex,costa2007characterization,bianconi2009assessing,shen2014collective}. Nodes represent entities in these systems and directed/undirected and weighted/unweighted links express relations among entities. It has been found that networks extracted from different fields possess some common features, such as power-law degree distribution, community structures and so on. The study of complex networks has become a substantial area of multidisciplinary research involving informatics, social sciences, physics, biology, mathematics, and other applied sciences.

In many cases, the data sets collected from various systems just reflect a part of reality. That means some links in observed networks are missing or may be able to appear in the future. This problem corresponds to one of the most critical tasks in network analysis, i.e. link prediction, which aims at estimating the likelihood of the existence of a link between a pair of nodes based on available network information \cite{liben2007link,lu2011link}. Predicting missing links has great value in many different applications. Sometimes, people concern latent links that are most likely to exist from the perspective of global network. For example, in the field of biology, as our knowledge of many biological systems is very limited, using predicted top-L links to guide the laboratory experiments rather than blindly checking all possible interactions will greatly reduce the experimental costs. While in some other cases, we want to predict or recommend personalized latent links for different nodes. For instance, social networking sites need to recommend new friends for different users to support better user experience or boost users' activity.

Link prediction in complex networks is an open issue and studied by researchers from disparate scientific communities in recent years. To solve this problem, two kinds of information is usually employed, including property and topological structure information. Property information can be utilized by many machine learning algorithms and very good results can be achieved \cite{al2006link,lichtenwalter2010new}. However, in some cases, property information is hard to be accessed because of privacy issue or reliability of collected data. Therefore, a kind of methods focus on solely topological information. Some other methods also consider the temporal uncertainty property of dynamic social networks \cite{ahmed2016efficient}.

Similarity-based framework is one of the most popular methods to solve link prediction problem \cite{liben2007link,lu2011link}. In this framework, each predicted pair of nodes is assigned a score to estimate the similarity between two nodes. A bunch of this kind of approaches has been developed, including methods based on local or global structure information.

In the early stage, some very simple similarity indices are used to predict missing links. PA index \cite{newman2001clustering} is defined as the product degrees of two nodes. CN index \cite{lorrain1971structural} counts the number of common-neighbors and JC \cite{jaccard1901distribution} is the normalization of CN. AA \cite{adamic2003friends} and RA \cite{zhou2009predicting} improve the resolution of CN and achieve better prediction accuracy. Recently, researchers proposed some more advanced common-neighbor based methods. Combining the idea of link community, Cannistraci et al. designed a series of CAR-based similarity indices to improve classical similarity indices with links between common-neighbors \cite{cannistraci2013link}. Some recent literatures show that CRA index, which is a variant of RA index, always performs best in all CAR-based indices. However, CAR-based methods are not efficient enough and need to be implemented with parallel technics to solve large networks. 

For higher accuracy, some global and quasi-local methods are proposed, such as Kaza \cite{katz1953new}, SimRank \cite{jeh2002simrank}, Hitting Time \cite{gobel1974random}, Average Commute Time \cite{fouss2007random}, Local Path \cite{lu2009similarity}, Transferring Similarity \cite{sun2009information}, Matrix Forest Index \cite{chebotarev1997matrix} and so on. Utilizing more information, these methods achieve better prediction results, but also cost more computation time. Besides, there are also some sophisticated methods, such as HSM \cite{clauset2008hierarchical} and SBM \cite{guimera2009missing}, which employ maximum likelihood estimation technics. L\"{u} et al proposed a structural perturbation method and concept of structural consistency, which can be used to investigate the link predictability \cite{lu2015toward} of networks. Pan et al proposed an algorithmic framework \cite{pan2016likelihood}, where a network¡¯s probability is estimated according to a predefined structural Hamiltonian, and the existence score of a non-observed link is quantified by the conditional probability of adding the focal link to the network while the spurious probability of an observed link is quantified by the conditional probability of deleting the link.

The study of link prediction is also closely related to the problem of network evolving mechanisms \cite{wang2012evaluating,zhang2015Measuring}. A recently proposed triangle growing network model can generate networks with various key features that can be observed in most real-world networks, with only two kinds of triangle growth mechanisms \cite{wu2015Emergence}. It indicates triangle information has crucial importance in link formation. A network index, called node clustering coefficient, is defined by the number of triangles passing through a node divided by the possible maximal number of triangles. And it has been proven node clustering information has good effectiveness in predicting missing links even under a very simple common-neighbor based framework \cite{wu2015efficient}.

CCLP index \cite{wu2015efficient} estimates the contribution of common-neighbors in forming a link between a pair of nodes by node clustering coefficient directly, and achieves good prediction results. Based on Bayes theory, LNBCN \cite{liu2011link} utilizes a local Na\"{i}ve Bayes model to differentiate the role of neighboring nodes. The final form of LNBCN index is a function of clustering coefficient plus a function of network density. Mutual information (MI) index \cite{tan2014link} is defined as the conditional self-information of the event that a pair of nodes are connected when their common-neighbors are given. In MI index, clustering coefficient can estimate the conditional probability of the existence of a link between a pair of nodes belonging to the ego-network of a given common-neighbor. From the final equations of these indices, node clustering coefficient always plays a key role. However, node clustering coefficient cannot distinguish different clustering abilities of a node to different node pairs. Unfortunately, real situations always dissatisfy this assumption. One node may play very different roles in forming different links in different circles. This is the main problem that we want to solve in this paper.

Besides, motif structures (can be considered as an extension from clustering) in directed networks are also explored in solving link prediction problem. Zhang et al proposed a hypothesis named potential theory, which assumes that every directed link corresponds to a decrease of a unit potential and subgraphs with definable potential values for all nodes are preferred. Combining the potential theory with the clustering and homophily mechanisms, it is deduced that the Bi-fan structure consisting of 4 nodes and 4 directed links is the most favored local structure in directed networks \cite{zhang2013Potential}. Their simulation results demonstrated that clustering (motif) information has critical importance for link prediction even in directed networks.

In this paper, we propose and investigate the power of local asymmetric clustering information to improve globalized and personalized top-L link prediction accuracy. As mentioned above, node clustering coefficient has some drawbacks in estimating the contribution of common neighbor nodes. Therefore, we shift our focus from node clustering to link clustering information. To different end nodes of a link, we define asymmetric link clustering coefficient, and our further studies show that the asymmetric link clustering information can be employed to predict top-L latent links with better performance than node clustering information.

The rest of this paper is organized as follows: Section \ref{sec:Definition} gives some fundamental definitions and concept of asymmetric link clustering. Section \ref{sec:Methods} presents three modified asymmetric link clustering information based link prediction methods. Section \ref{sec:Experiments} demonstrates the experimental results. Section \ref{sec:Conclusions} offers conclusions and related prospects in the future work. 

\section{DEFINITION}
\label{sec:Definition}
Definition 1. Graph- A network can be represented by a graph $G = (V, E)$, where $V$ is the set of nodes and $E$ is the set of links. $\widetilde{E}$ is a set representing all non-existent links, i.e. unconnected node pairs.

Definition 2. Latent links - Each node-pair in $\widetilde{E}$ will be assigned a similarity score and sorted in descend order. The latent links at the top of the rank are most likely to exist.

Definition 3. Node Clustering (NC) coefficient- The clustering coefficient of a node is defined as the number of triangles passing through a node divided by possible maximum number of triangles passing through this node. It can be calculated by equation (\ref{eq:Cz}).

\begin{equation}
C_z = \frac{t_z}{k_z(k_z-1)/2}
\label{eq:Cz}
\end{equation}
where $t_z$ is the number of triangles passing through node $z$, and $k_z$ is the degree of node $z$.

Definition 4. Asymmetric link clustering (ALC) coefficient- The motivation of link clustering coefficient is similar with that of node clustering coefficient. In general, it can be defined as the number of triangles passing through a link divided by possible maximum number of triangles \cite{wang2012identification}. However, in link prediction problem, we always have more interests in the possible maximum number of triangles related to one of the two end nodes, because different end-nodes may play different roles. For example, one end node may belong to the predicted pair of nodes, and the other one may belong to common-neighbors of the predicted pair of nodes. Therefore, we propose an asymmetric link clustering coefficient defined by equation (\ref{eq:lc}).

\begin{equation}
LC_{x,z} = \frac{|\Gamma(x) \cap \Gamma(z)|}{k_z-1}
\label{eq:lc}
\end{equation}
where $\Gamma(x)$ denotes the set of neighbors of node $x$ and $|A|$ is the number of items in set A.

It is obvious that $LC_{x,z}$ is not equal to $LC_{z,x}$, unless the degree of node $x$ is equal to that of node $z$. The asymmetric link clustering coefficient measures the clustering ability of a link from the perspective of each of its two end-nodes. In this work, we will investigate the crucial importance to differentiate the clustering ability of a link from the perspective of its two end-nodes in predicting missing links.

In this paper, several classical and advanced local link prediction methods are compared and analyzed. Their definitions are given as follows:

(1)	CN (Common Neighbors) index \cite{lorrain1971structural} counts the number of common neighbors for a pair of nodes. In general, more common neighbors indicate larger probability to form/exist a link between two nodes. The definition of CN is given in equation (\ref{eq:cn}).

\begin{equation}
s^{CN}_{xy} = |\Gamma(x) \cap \Gamma(y)|
\label{eq:cn}
\end{equation}

(2)	LocalPath index \cite{lu2009similarity} considers the information of the next nearest neighbors, which is a kind of extension of CN index.

\begin{equation}
s^{LocalPath}_{xy} = (A^2)_{xy} +\varepsilon(A^3)_{xy}
\label{eq:lp}
\end{equation}
where $\varepsilon$ is a parameter. $(A^2)_{xy}$ and $(A^3)_{xy}$ are the numbers of different paths with length 2 and 3 connecting $x$ and $y$, respectively. In our experiments, $\varepsilon$ is fixed as 0.01, which is recommended by its authors.

(3)	RA (Resource Allocation) \cite{zhou2009predicting} is an index based on common neighbors, and the motivation comes from resource allocation dynamics on complex systems. For a pair of unconnected nodes, $x$ and $y$, the node $x$ can send some resource to $y$, with their common neighbors playing the role of transmitters. In the simplest case, assume that each transmitter has a unit of resource, and will equally distribute the resource to all its neighbors. 

\begin{equation}
s^{RA}_{xy} = \sum_{z\in\Gamma(x)\cap\Gamma(y)}\frac{1}{k_{z}}
\label{eq:ra}
\end{equation}

(4)	CRA index \cite{cannistraci2013link} is a revised version of RA index based on link community paradigm, which includes some additional consideration on the link structure between common-neighbors. The definition of CRA is

\begin{equation}
s^{CRA}_{xy} = \sum_{z\in\Gamma(x)\cap\Gamma(y)}\frac{|\gamma(z)|}{|\Gamma(z)|}
\label{eq:cra}
\end{equation}
where $\gamma(z)$ refers to the subset of neighbors of node $z$ that are also common neighbors of nodes $x$ and $y$.

(5)	CCLP index \cite{wu2015efficient} utilizes node clustering coefficient (equation (\ref{eq:Cz})) to estimate the contribution of common neighbors directly. Its definition is in equation (\ref{eq:cclp}).

\begin{equation}
s^{CCLP}_{xy} = \sum_{z\in\Gamma(x)\cap\Gamma(y)}C_z
\label{eq:cclp}
\end{equation}

(6)	LNBCN index \cite{liu2011link} is inferred according to the Bays theory. The final definition of LNBCN is given in equation (\ref{eq:lnbcn}). It consists two parts: the first part is a function of node clustering coefficient and the second part is a function of network density.

\begin{equation}
s^{LNBCN}_{xy} =\sum_{z\in\Gamma(x)\cap\Gamma(y)}(log(\frac{C_z}{1-C_z})+log(\frac{1-\rho}{\rho}))
\label{eq:lnbcn}
\end{equation}
where $\rho$ is the network density and is defined in equation (\ref{eq:rho}).

\begin{equation}
\rho = \frac{|E|}{|V|(|V|-1)/2}
\label{eq:rho}
\end{equation}

(7)	MI (Mutual Information) index \cite{tan2014link} is defined based on information theory as shown in equation (\ref{eq:mi}). If we utilize network density to infer self-information $I(A^1_{xy})$ and node clustering coefficient to infer conditional self-information $I(A^1_{xy}|z)$, then the mutual information $I(A^1_{xy};z)$ can be represented as equation (\ref{eq:mi_cal}).

\begin{equation}
s^{MI}_{xy} = \sum_{z\in\Gamma(x)\cap\Gamma(y)}I(A^1_{xy};z)-I(A^1_{xy})
\label{eq:mi}
\end{equation}

\begin{equation}
I(A^1_{xy};z) = I(A^1_{xy})-I(A^1_{xy}|z)=-logP(\rho)+logP(C_z);
\label{eq:mi_cal}
\end{equation}

CN index is the most basic common-neighbor based method. LocalPath index improves CN from the perspective of considering longer path information, while RA and CCLP improve CN from the perspective of distinguishing the function of different common-neighbors. CRA further improves RA with more local link structure information. LNBCN and MI consider the link prediction problem under different theories, while the final forms also belong to common-neighbor based local methods.

\section{Methods}
\label{sec:Methods}
From the three last introduced link prediction methods, i.e. CCLP, LNBCN and MI, we can observe that clustering information in solving link prediction problem has crucial importance. Although the three methods start from different perspectives, clustering coefficient plays important role in the final definitions. However, node clustering coefficient utilized by these methods is universally invariant. That means these methods consider each node playing the same role for predicting different latent links in different local networks. Actually, a node can belong to different circles or communities and play different roles. Ahn et al. has proposed the concept of link community several years ago \cite{ahn2010link}. They found that starting from the view of links information that is more diverse could be achieved, such as overlapping community structures.

Therefore, we focus on the power of asymmetric link clustering information in predicting globalized and personalized top-L missing links. The definition of asymmetric link clustering coefficient has been given in equation (\ref{eq:lc}). The so-called asymmetry means that the clustering coefficient of a link can be different when its two end-nodes play different roles. Based on this idea, we will give three modified clustering based link prediction methods under three different frameworks in the following subsections.

\subsection{Asymmetric topological structure based method}
\label{subsec:topological}
In this subsection, we present a simple similarity index based on asymmetric link clustering information under the framework of common-neighbors. In this method, the similarity is first estimated from the perspective of two different seed-nodes, respectively. Here two seed-nodes correspond to a predicted pair of nodes. 

\begin{figure*}[ht]
\centering
\includegraphics[width=12cm]{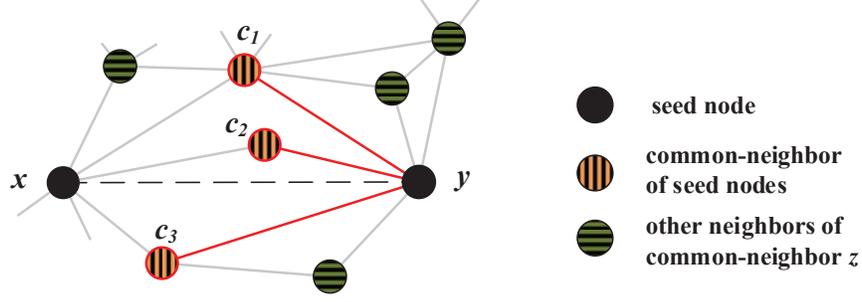}
\caption{Illustration of asymmetric link clustering information based similarity index.}
\label{fig:illustration}
\end{figure*}

As shown in Fig. \ref{fig:illustration}, there are three common-neighbors between seed-node x and y. For each seed-node, there are three links connecting common-neighbors, respectively. The new method, called ACC, focuses on the clustering information of these links existing between seed-nodes and common-neighbors. A larger link clustering coefficient indicates larger probability of forming a triangle or existing a link between seed nodes. The similarity can be defined as the summation of asymmetric link clustering coefficient of links connecting one seed node and common-neighbors.

It can be noted that the similarity of node pair $(x, y)$ can be estimated from the view of either node $x$ or node $y$. For example, we can define the similarity of $(y, x)$ by equation (\ref{eq:accPre}). In this equation, only clustering information of links, i.e. the red heavy links in Fig. \ref{fig:illustration}, is used.

\begin{equation}
sim(y,x) = \sum_{z\in\Gamma(x)\cap\Gamma(y)}LC_{y,z}
\label{eq:accPre}
\end{equation}

It is obvious that $sim(x, y)$ is not equal to $sim(y, x)$ from the definition of equation (\ref{eq:accPre}). Thus, when choosing top-L latent links in predicting procedure, we can rank all asymmetric node pairs according to their similarities. Since we only predict undirected links, either $sim(x, y)$ or $sim(y, x)$ is ranked in top-L, the undirected node pair $(x, y)$ is judged as a missing link. In other words, the result is only dominated by the higher value between $sim(x, y)$ and $sim(y, x)$. Therefore, the final ACC similarity index only chooses the higher similarity. The definition is given in equation (\ref{eq:acc}).

\begin{equation}
s^{ACC}_{xy} = max\{\sum_{z\in\Gamma(x)\cap\Gamma(y)}LC_{x,z}, \sum_{z\in\Gamma(x)\cap\Gamma(y)}LC_{y,z}\}
\label{eq:acc}
\end{equation}

\subsection{Asymmetric Local Na\"{i}ve Bays based method}
\label{subsec:Bays}
Local Na\"{i}ve Bays based model is a probability-based method \cite{liu2011link}. As shown in equation (\ref{eq:lnbPre}), the key problem of this method is to define the probability of existing a link or not for a given sub-network regarding to a pair of seed-nodes, properly.

\begin{equation}
sim(x,y) = \frac{P(A^1_{xy}|G_{xy})}{P(A^0_{xy}|G_{xy})}
\label{eq:lnbPre}
\end{equation}
where $A^1_{xy}$ and $A^0_{xy}$ indicate the event of existing a link or not between node $x$ and $y$, respectively. $G_{xy}$ indicates the sub-network regarding to $x$ and $y$. In our method, we continue to solve this problem utilizing asymmetric link clustering information. Thus, the two conditional probabilities are defined as equation (\ref{eq:PA1xy}) and (\ref{eq:PA0xy}). Here $L_x = \{e_{xz}|~ e_{xz} \in E~,~z \in \Gamma(x)\cap\Gamma(y)\}$.

\begin{equation}
P(A^1_{xy}|L_x) = \frac{P(A^1_{xy})}{P(L_x)}\prod_{e\in L_x} P(e|A^1_{xy})
\label{eq:PA1xy}
\end{equation}

\begin{equation}
P(A^0_{xy}|L_x) = \frac{P(A^0_{xy})}{P(L_x)}\prod_{e\in L_x} P(e|A^0_{xy})
\label{eq:PA0xy}
\end{equation}

Substituting equation (\ref{eq:PA1xy}) and (\ref{eq:PA0xy}), we can get

\begin{equation}
sim(x,y) = \frac{P(A^1_{xy})}{P(A^0_{xy})}\prod_{z\in \Gamma(x)\cap\Gamma(y)}\frac{P(A^0_{xy})\cdotp P(A^1_{xy}|e_{xz})}{P(A^1_{xy})\cdotp P(A^0_{xy}|e_{xz})}
\label{eq:lnbPre2}
\end{equation}

$P(A^1_{xy})$ can be estimated by the whole network density as defined in equation (\ref{eq:rho}), and $P(A^0_{xy})$ is equal to $1-P(A^1_{xy})$. Since $P(A^1_{xy})/P(A^0_{xy})$ is the same for all pairs of nodes, it does not affect the rank of latent links and can be removed in the final definition of similarity index. $P(A^1_{xy}|e_{xz})$ can be estimated by asymmetric link clustering coefficient $LC_{xz}$, and $P(A^0_{xy}|e_{xz})$ is equal to $1-P(A^1_{xy}|e_{xz})$.

At last, the situation is the same with ACC. $sim(x,y)$ is not equal to $sim(y,x)$, but the results are only influenced by the higher value. Equation (\ref{eq:alnb}) shows the final form of ALNB index.

\begin{equation}
\begin{split}
s^{ALNB}_{xy} = max\{\prod_{z\in\Gamma(x) \cap \Gamma(y)}\frac{P(A^0_{xy})\cdotp P(A^1_{xy}|e_{xz})}{P(A^1_{xy})\cdotp P(A^0_{xy}|e_{xz})}, \\
\prod_{z\in\Gamma(x) \cap \Gamma(y)}\frac{P(A^0_{xy})\cdotp P(A^1_{xy}|e_{yz})}{P(A^1_{xy})\cdotp P(A^0_{xy}|e_{yz})}\}
\label{eq:alnb}
\end{split}
\end{equation}

\subsection{Asymmetric Mutual Information based method}
\label{subsec:mutual}
In this subsection, we modify the mutual information based method via the concept of ALC. First, we introduce the definition of self-information and mutual information. In information theory, self-information is defined in equation (\ref{eq:si}), and the conditional self-information can be calculated by self-information and mutual information as shown in equation (\ref{eq:mi2}).

\begin{equation}
I(x) = -log P(x)
\label{eq:si}
\end{equation}

\begin{equation}
I(x|y) = I(x) - I(x;y)
\label{eq:mi2}
\end{equation}

Mutual information based method defines the similarity as a conditional self-information, as shown in equation (\ref{eq:miPre}) \cite{tan2014link}. Here, we still consider $L_x = \{e_{xz}|~ e_{xz} \in E~,~z \in \Gamma(x)\cap\Gamma(y)\}$ as the part of $G_{xy}$.

\begin{equation}
sim(x,y) = - I(A^1_{xy}|G_{xy})=I(A^1_{xy};L_x)-I(A^1_{xy})
\label{eq:miPre}
\end{equation}

Here $P(A^1_{xy})$ can be estimated by the whole network density as defined in equation (\ref{eq:rho}). In Ref. ~\cite{tan2014link}, this probability is defined by a more complex method, but we find these two kinds of definitions of $P(A^1_{xy})$ only perform better than each other on different networks. The authors of MI index also recommended this method in their another work \cite{zhu2015information}. Therefore, the part of $I(A_{xy}^1)$ can be removed, because it is the same for all pairs of nodes, and does not affect the final rank of latent links. Thus, equation (\ref{eq:miPre}) can be simplified as equation (\ref{eq:miPre2}).

\begin{equation}
sim(x,y) = \sum_{z\in\Gamma(x)\cap\Gamma(y)}I(A^1_{xy};e_{xz})
\label{eq:miPre2}
\end{equation}

Then we try to calculate the above mutual information. According to equation (\ref{eq:mi2}), $I(A_{xy}^1;e_{xz})$ can be represented as 

\begin{equation}
I(A^1_{xy};e_{xz}) = I(A^1_{xy}) - I(A^1_{xy}|e_{xz})
\label{eq:mi3}
\end{equation}

Thus, we need to estimate the probability $P(A_{xy}^1 |e_{xz})$, which can be calculated by the link clustering coefficient $LC_{xz}$. At last, we still choose higher value between $sim(x, y)$ and $sim(y, x)$ to form the final definition of AMI index, which is given by equation (\ref{eq:ami}).

\begin{equation}
s^{AMI}_{xy} = max\{\sum_{z\in\Gamma(x) \cap \Gamma(y)}I(A^1_{xy};e_{xz}), \sum_{z\in\Gamma(x) \cap \Gamma(y)}I(A^1_{xy};e_{yz})\}
\label{eq:ami}
\end{equation}

\section{Experiments}
\label{sec:Experiments}
In this section, we will give the experimental results of the proposed asymmetric link clustering (ALC) based link prediction methods on 12 networks collected from various fields. Seven neighborhood based link prediction methods are compared, including three corresponding NC based methods and four other representative methods. Please note in most related literatures, authors only give the results of their methods in solving globalized link prediction problem, here we demonstrate the performance of these methods in predicting both globalized and personalized top-L latent links.

\subsection{Data sets}
\label{subsec:Data}
In the following experiments, we will test our methods on 6 small networks (the number of nodes is less than 1000) and 6 large networks (the number of nodes is more than 1000) from social, biological and technological fields. Except local methods, we also compare our methods with 2 global link prediction methods, i.e. HSM \cite{clauset2008hierarchical} and SBM \cite{guimera2009missing} on 6 small networks. Dolphins \cite{lusseau2003bottlenose} is a dolphin social network, in which an edge represents frequent association between a pair of dolphins. Mouse \cite{bock2011network} is a mouse neural network and the synaptic connections between neurons are represented as links. In Macaque \cite{kotter2004online} network, cortical connectomes are modeled as nodes and links. Food \cite{ulanowicz1998network} is a food web network. Celegans \cite{watts1998collective} is a neural network of the nematode Caenorhabditis elegans. Yeast \cite{bu2003topological} is a protein-protein interaction network of yeast. Hamster \cite{hamster2015} is a friendship network between users of the website hamsterster.com. PB \cite{adamic2005political} is a political blog network. USAir \cite{usair} is a network of US air transportation system. INT \cite{spring2002measuring} is a network of router-level topology of the Internet. P2P \cite{leskovec2007graph} is a peer-to-peer file sharing network. Scimet \cite{scimet} is a network of articles from or citing Scientometrics. The statistics of these networks are given in Table \ref{tab:netStatics}. Directed links are treated as undirected ones, and self-loops are removed.

\begin{table}[ht]
\centering
\begin{tabular}{|l|l|l|l|l|l|l|l|l|}
\hline
Nets        &$|V|$      &$|E|$     &$\langle d\rangle$     &$\langle k\rangle$     &$H$ & $\langle C\rangle$ & $\langle LC\rangle$ & $r$ \\  
\hline 
Dolphins&62&159&3.357&5.129&1.327&0.259&0.118&-0.044\\ 
Mouse&18&37&1.967&4.111&1.289&0.216&0.108&-0.516\\ 
Macaque&94&1515&1.771&32.234&1.238&0.774&0.585&-0.151\\ 
Food&128&2075&1.776&32.422&1.237&0.335&0.113&-0.112\\ 
Celegans&297&2148&2.455&14.465&1.801&0.292&0.081&-0.163\\ 
USAir&332&2126&2.738&12.807&3.464&0.625&0.37&-0.208\\
Hamster&1788&12476&3.453&13.955&3.264&0.143&0.03&-0.085\\ 
PB&1222&16714&2.738&27.355&2.971&0.32&0.103&-0.221\\ 
Yeast&2375&11693&5.096&9.847&3.476&0.306&0.273&0.454\\ 
INT&5022&6258&6.449&2.492&5.503&0.012&0.009&-0.138\\ 
P2P&6299&20776&4.642&6.597&2.676&0.011&0.002&0.036\\ 
Scimet&2678&10368&4.18&7.743&2.427&0.174&0.049&-0.033\\ 
\hline
\end{tabular}
\caption{\label{tab:netStatics}The basic topological features of twelve real world networks. $|V|$ and $|E|$ are the number of nodes and links. $\langle d\rangle$ is average shortest distance. $\langle k\rangle$ is average degree. $H$ is the degree heterogeneity, defined as $H=\frac{\langle k^2\rangle}{\langle k^2\rangle}$. $\langle C\rangle$ and $\langle LC\rangle$ are average node and link clustering coefficient, and $r$ is the assortative coefficient \cite{newman2002assortative}.}
\end{table}

\subsection{Evaluation}
\label{subsec:Evaluation}
To evaluate the performance of compared methods in predicting top-L latent links, we employ two estimators: precision and AUP. The preparations for the two chosen estimators are the same. First, the link set $E$ of a network is randomly divided into two parts: $E_t$ and $E_p$. The training set $E_t$ is used as the known topological information, and the probe set $E_p$ is used as the test set. We can get $E_t \cup E_p = E$ and $E_t \cap E_p = null$. To make sure that the training set contains true structure information, we only set 10\% of links as the test links, because removing too many links from the network may destroy the structure of the original network.

We employ a commonly used way to estimate the effectiveness of the top-L latent link prediction methods. First we calculate the similarity of all unconnected node pairs, including those belonging to probe set $E_p$ and nonexistent link set $\widetilde{E}$. Then all these latent links are ranked in descend order.

Definition 5. Precision- The precision is defined as the ratio of relevant items selected to the number of items selected \cite{herlocker2004evaluating}. That means if we take top $L$ links as the predicted ones, among which $L_r$ links are right, then the precision can be defined as equation (\ref{eq:prec}). Higher precision indicates higher prediction accuracy. In globalized link prediction, $L$ is set to 20 or 100 for networks with less or more than 1,000 links. In personalized link prediction, $L$ is set to 5, and the final precision is the average of precisions of all nodes.

\begin{equation}
precision = \frac{L_r}{L}
\label{eq:prec}
\end{equation}

Definition 6. AUP- AUP is defined as the Area Under Precision curve \cite{cannistraci2013link}. Here, precision curve is achieved by using ten different values of $L$, which are 10,20,...,100 for networks with more than 1000 links and 2, 4,...,20 for networks with less than 1000 links in globalized link prediction and are 1, 2,...,5 in personalized link prediction.

\subsection{Globalized top-L latent link prediction}
\label{subsec:Globalized}
In this subsection, we give the experimental results of these methods in predicting globalized top-L latent links. The so-called "globalized" here means the top-L latent links are selected from the perspective of the whole network. For example, in biology experiments, we need to find some most probable candidate interactions between proteins to check in the whole system. 

\begin{figure*}
\centering
\includegraphics[width=13cm]{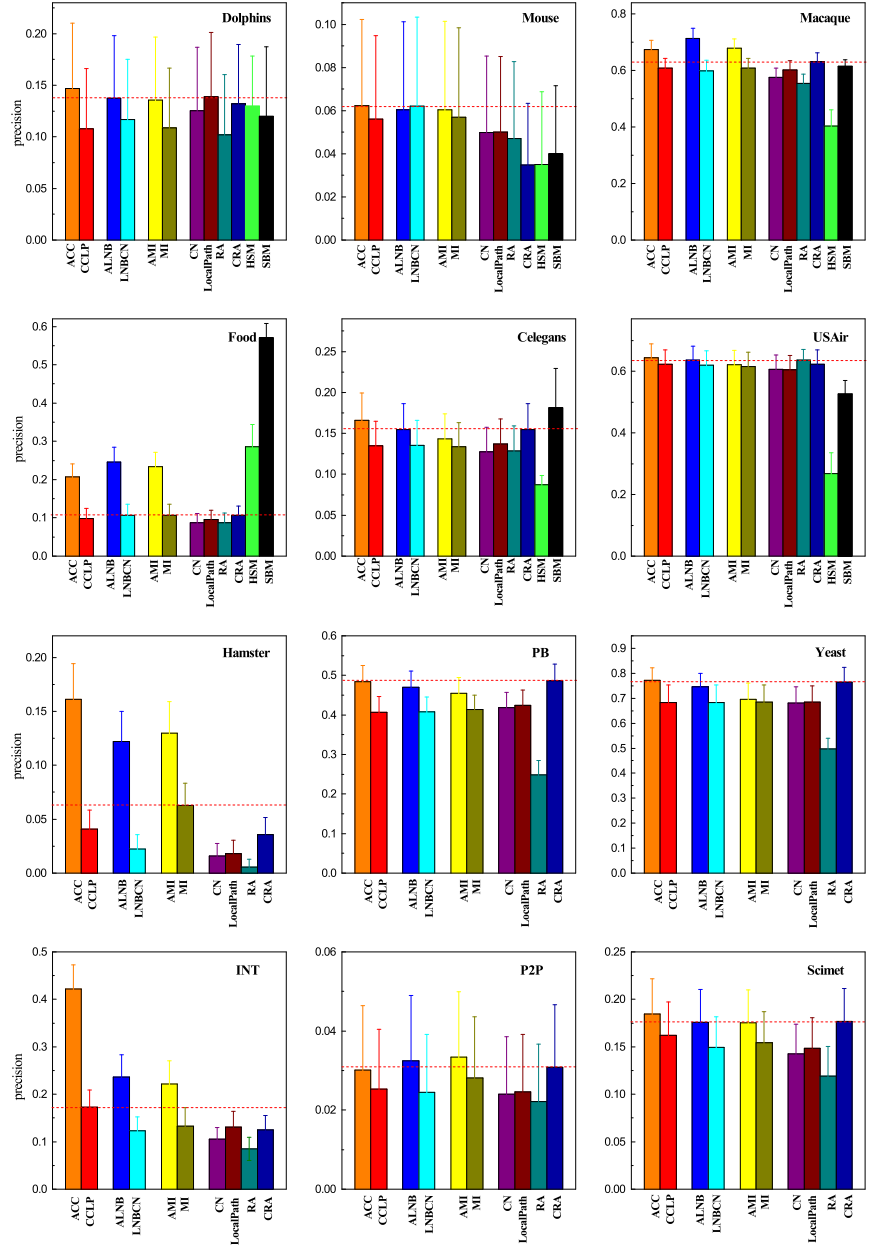}
\caption{Globalized Link prediction accuracy in 12 networks estimated by precision.}
\label{fig:globalPrecision}
\end{figure*}

Figure \ref{fig:globalPrecision} shows the globalized link precision results of these methods on 12 networks. On the top six small networks, 12 link prediction methods are compared, including two global methods, i.e. HSM and SBM. It can be noted that asymmetric link clustering information based methods perform better than their corresponding node clustering based methods in most cases, and the lowest accuracy attained by asymmetric link clustering based methods is even always higher than the highest accuracy attained by other node clustering based methods. Comparing with global methods, ALC-based methods achieve better results on four of six networks. Only on Food network, global methods demonstrate absolute advantage, while on other small networks, ALC-based methods do not perform worse than global methods significantly. 

In Fig. \ref{fig:globalPrecision}, the given straight line indicates the highest accuracy of all other non-ALC based local link prediction methods. In all 12 tested networks, ALC based methods almost achieve all best results among local methods. In eight networks, the best results of ALC based methods are significantly better than the highest accuracy of non-ALC based local methods. Especially for food web, Hamster and INT networks, the ratios improved by ALC-based methods come up to more than 100\% comparing with their counterparts. In other four networks, ALC based methods achieve similar precision with the best results of other seven compared local link prediction methods. 

\begin{figure*}
\centering
\includegraphics[width=12cm]{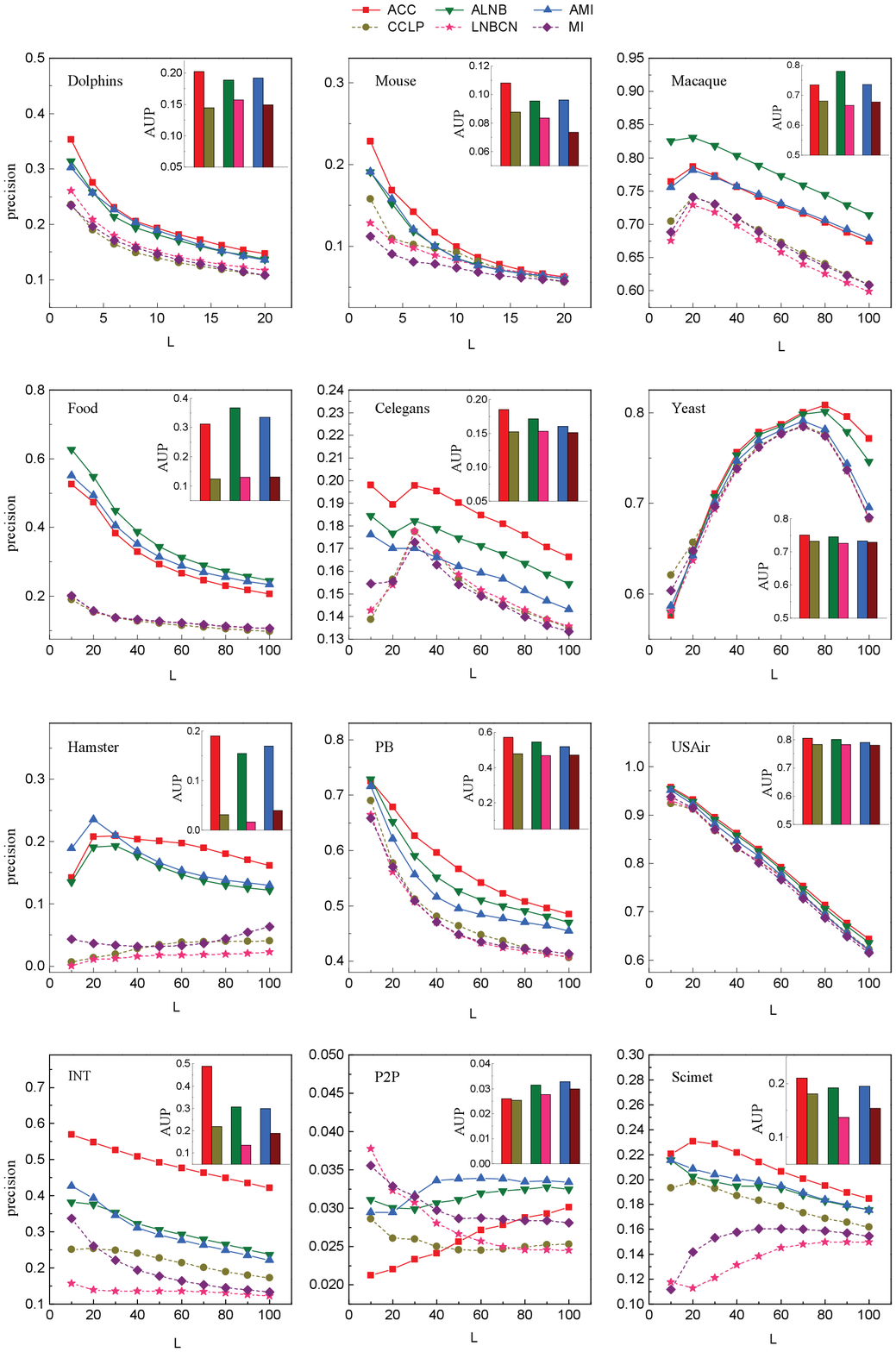}
\caption{Precision curves and AUP of ALC and their corresponding NC based methods for globalized link prediction. The columns of AUP, from left to right, are ACC, CCLP, ALNB, LNBCN, AMI and MI.}
\label{fig:globalPrecCurve}
\end{figure*}

To further investigate the dependence of globalized link prediction accuracy on the value of $L$, we plot the precision curves on each network with various values of $L$ and the corresponding AUP results in Fig. \ref{fig:globalPrecCurve}. To make the figure clearer, we only give the precision curve of three ALC based methods and their corresponding NC based methods. From the results in Fig. \ref{fig:globalPrecCurve}, we find that the change of $L$ does not affect the improvements made by asymmetric link clustering information in most networks.

\begin{figure*}
\centering
\includegraphics[width=12cm]{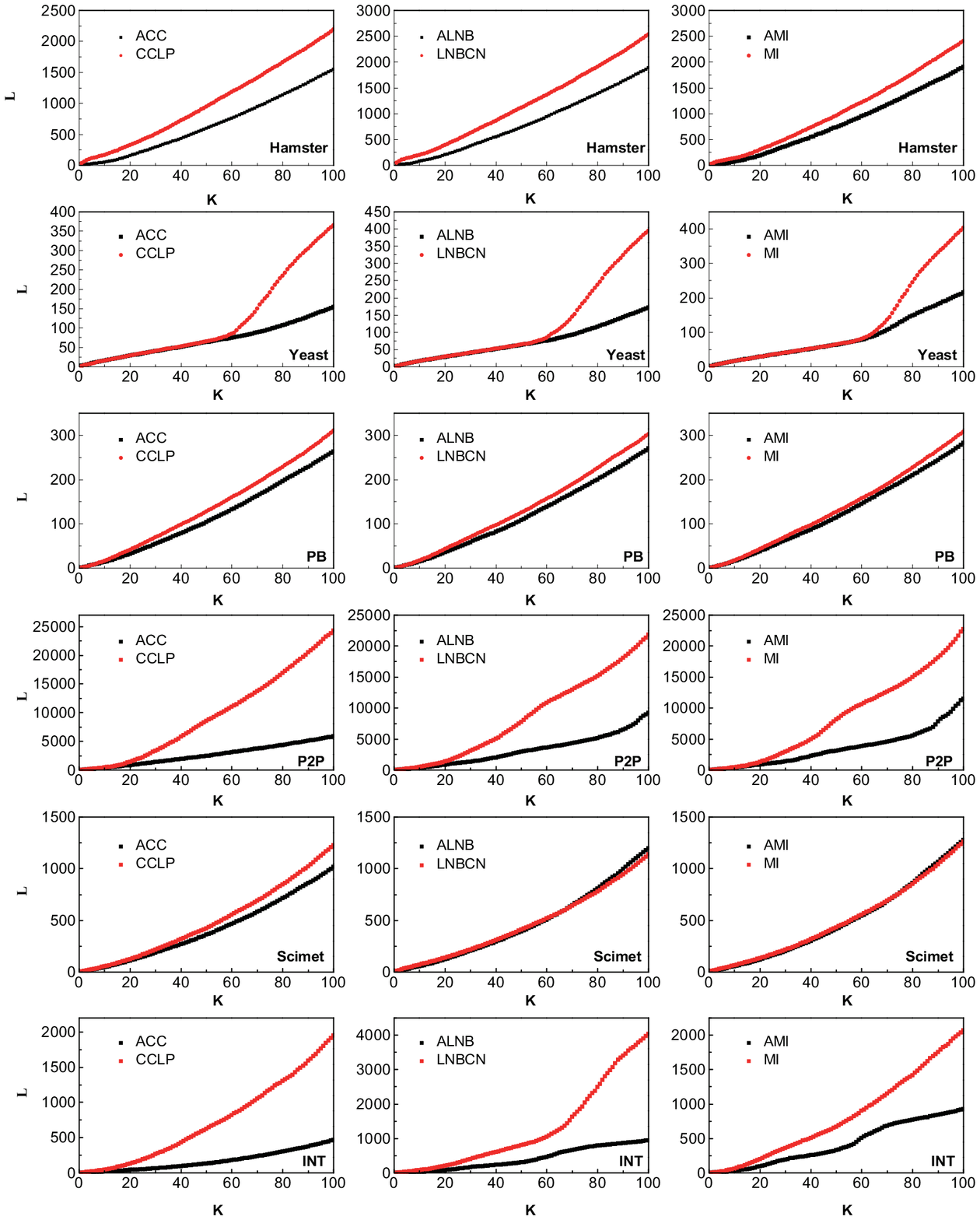}
\caption{The needed length of candidates ($L$) to hit $K$ links for ALC and NC based methods in globalized link prediction on six relatively large networks. The horizontal axis $K$ represents the number of latent links that an algorithm hits and the vertical axis $L$ presents the average number of node pairs we need to observe in the whole rank of node pairs.}
\label{fig:globalHitKL}
\end{figure*}

The above analysis only focuses on top-100 candidate links. To further check the power of ALC based methods, we investigate the needed length of candidates $L$ to hit $K$ links on six relatively large networks in Fig. \ref{fig:globalHitKL}. In this experiment, $K$ is set from 1 to 100, which means the observed scope of candidates is largely extended, and the differences among ALC and NC based methods are more notable. For a same value of $K$, a smaller $L$ indicates better performance.

Overall, the advantages of ALC based methods are very stable. With the increase of $K$, the improvements made by ALC based methods always become larger. For some networks, such as Yeast network, when $K$ is less than 60, there is almost no difference between the results of ALC and NC based methods. However, when $K$ is larger than 60, ALC based methods demonstrate remarkable advantages. This result indicates the value of parameter $L$ does not influence the advantage of ALC based methods, and a larger $L$ always makes the improvements even more remarkable in globalized top-L latent link prediction.

\subsection{Personalized top-L latent link prediction}
\label{subsec:Personalized}

In this subsection, we demonstrate the experimental results of these methods in personalized top-L latent link prediction. Personalized link prediction aims at predicting latent links for each node, respectively. For example, social networking sites always need to recommend new friends for their users. In this scenario, all users not connecting to a specific user will be sorted according their similarities with this node in descend order, and top $L$ selected users are recommended to this user. 

In the following analysis, the given accuracy is the average of predicting precision of all nodes. In our experiments, since the links in probe set $E_p$ are randomly selected, some nodes may have more links removed and some other nodes may have fewer missing links. Here the value of parameter $L$ just defines the maximal number of selected links. If the number of removed links is smaller than $L$, $L$ will be modified as the number of removed links for this node. Please note different strategies can be adopted here. We believe that as long as the adopted strategy is consistent, the results are fair to each compared method. Other configurations are the same with the experiments in globalized link prediction.

\begin{figure*}
\centering
\includegraphics[width=13cm]{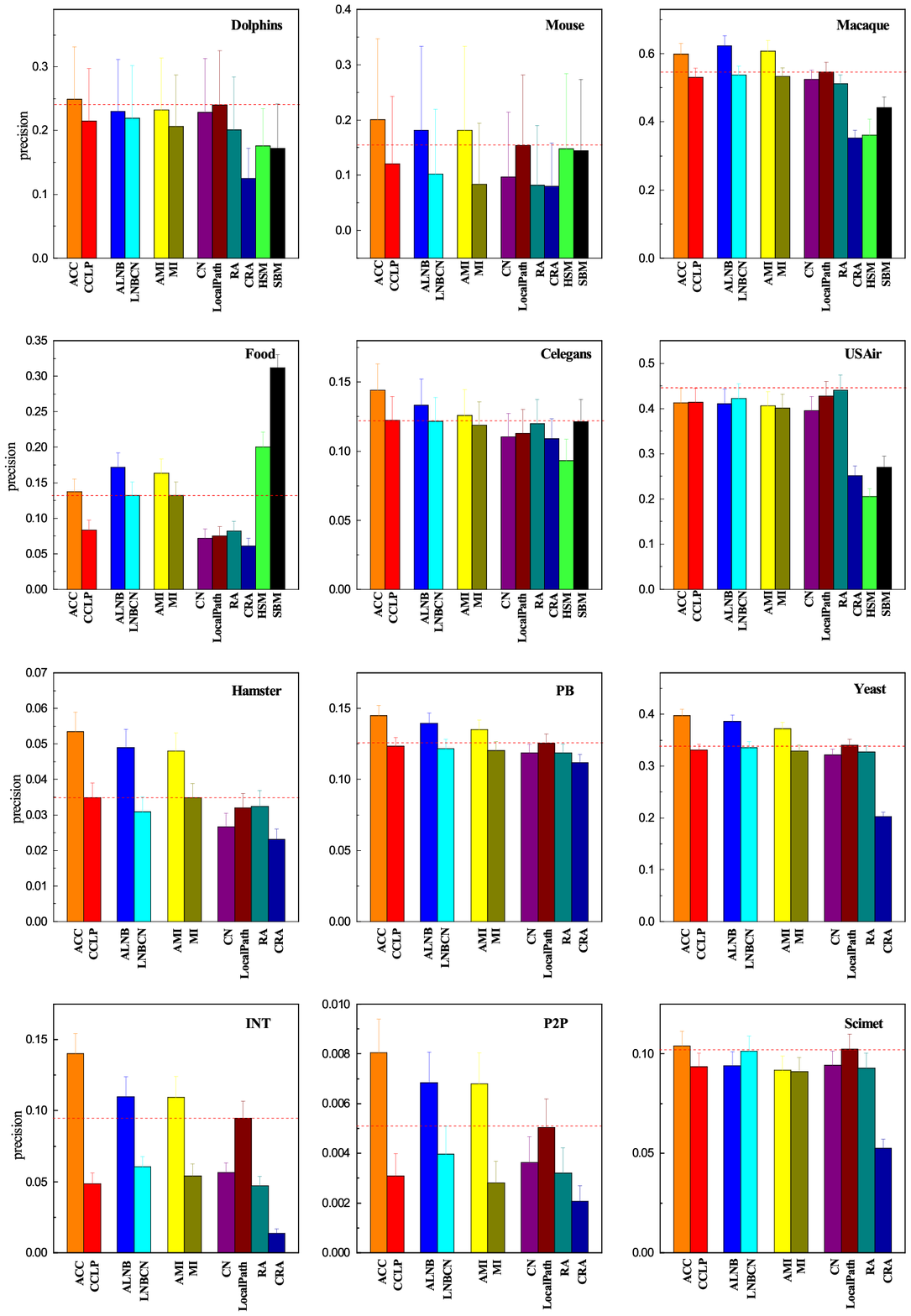}
\caption{Personalized Link prediction accuracy in 12 networks estimated by precision.}
\label{fig:personalPrecision}
\end{figure*}

Fig. \ref{fig:personalPrecision} demonstrates the personalized link prediction accuracy on 12 networks. On the top six small networks, we compare 12 methods, including HSM and SBM. The results are similar to those of globalized experiments, i.e. except on Food network, ALC-based methods do not perform worse than global methods on most networks. We can also observe clear improvements made by ALC based methods to their corresponding NC based methods once again in most networks. In most cases, the best results are also achieved by ALC based methods, except on USAir network, in which the classical RA index performs best. 

We can find another interesting phenomenon, i.e. for globalized link prediction, CRA index always performs very well in all non-ALC based methods, while for personalized link prediction, LocalPath index always performs better than other non-ALC based methods. The predicting accuracy of CRA index in personalized link prediction has a significant drop. However, the performance of ALC based methods are always stably well in both globalized and personalized latent link prediction. That demonstrates the advantage of asymmetric link clustering information in predicting missing links.

\begin{figure*}
\centering
\includegraphics[width=12cm]{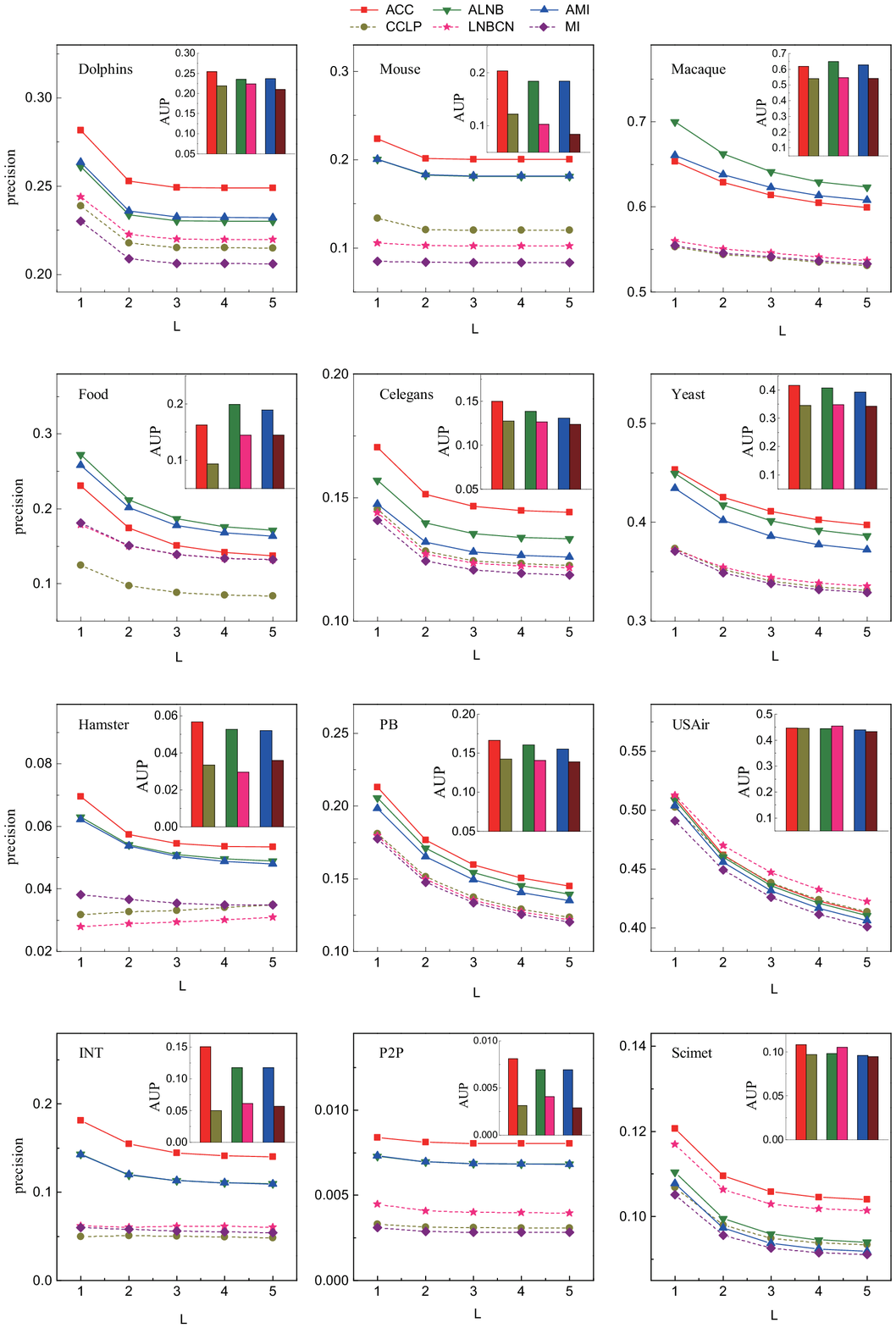}
\caption{Precision curves and AUP of ALC and their corresponding NC based methods for personalized link prediction. The columns of AUP, from left to right, are ACC, CCLP, ALNB, LNBCN, AMI and MI.}
\label{fig:personalPrecCurve}
\end{figure*}

\begin{figure*}
\centering
\includegraphics[width=12cm]{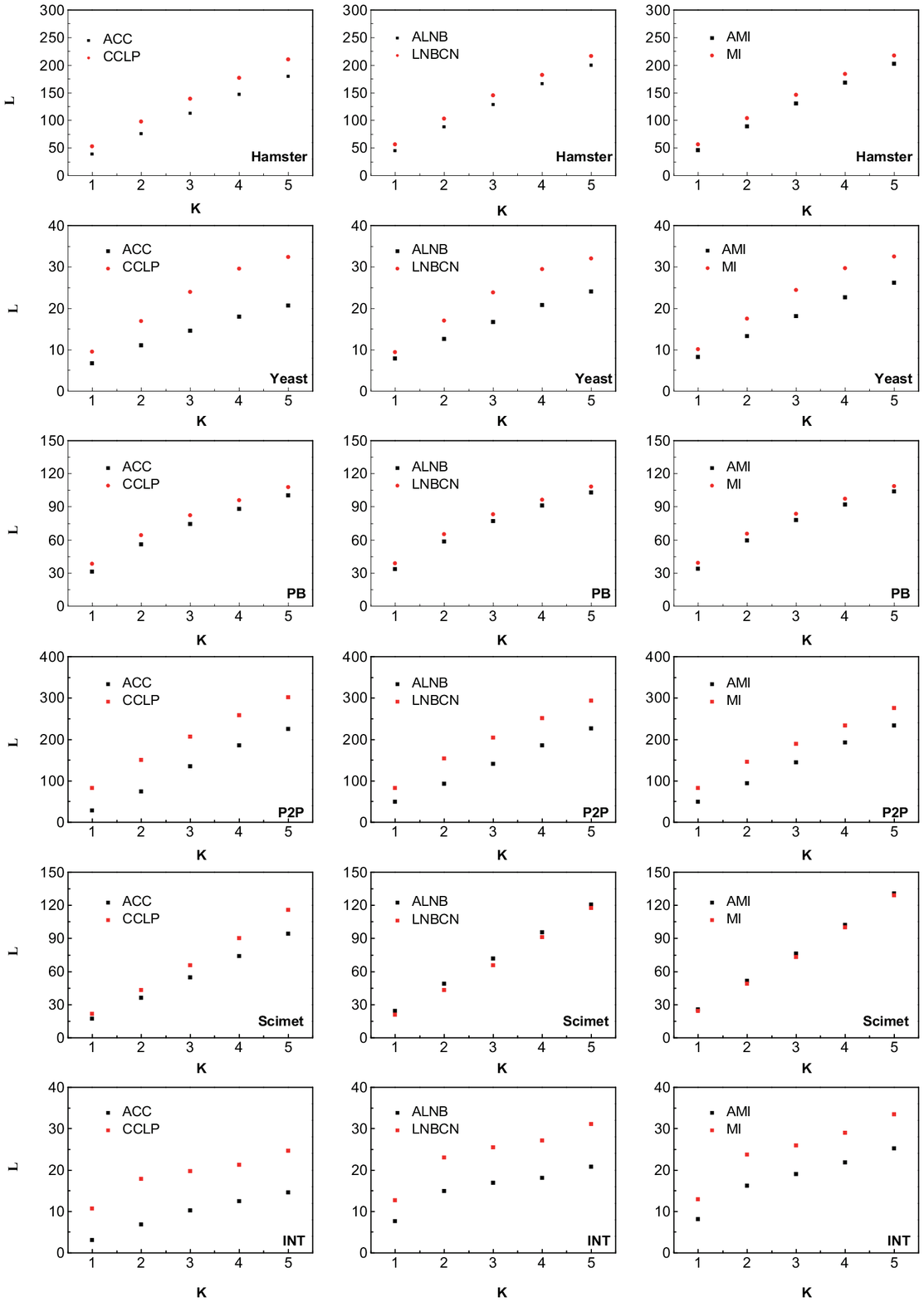}
\caption{The needed length of candidates ($L$) to hit $K$ links for ALC and NC based methods in personalized link prediction on six relatively large networks. The horizontal axis $K$ represents the number of latent links that an algorithm hits and the vertical axis $L$ presents the average number of node pairs we need to observe in the whole rank of node pairs.}
\label{fig:personalHitKL}
\end{figure*}

Fig. \ref{fig:personalPrecCurve} gives the precision curves and AUP with the value of L from 1 to 5 for personalized link prediction. The results are even clearer than those in globalized link prediction are. In most cases, all precision curves of ALC-based methods have clearly higher positions than corresponding NC based methods. In Fig. \ref{fig:personalHitKL}, we also give the needed length of $L$ to hit $K$ links in personalized link prediction. Here $K$ is set from 1 to 5. The results show that the performances of ALC based methods are very stable and not impacted by the chosen parameter $L$ in personalized latent link prediction.

\subsection{Time costs}
\label{subsec:Time}
Table \ref{tab:time} gives the time costs of all compared link prediction methods. Since global methods have some efficiency limitation, they are only tested on small networks. ALC based methods demonstrate competitive efficiency with NC based methods. Each result is an average of 100 independent runs.

\begin{table*}[ht]
\centering
\begin{tabular}{|l|l|l|l|l|l|l|l|l|l|l|l|l|}
\hline
Nets &ACC &CCLP &ALNB &LNBCN &AMI &MI &CN &LocalPath&RA&CRA&HSM&SBM \\  
\hline 
Dolphins&0.001 &0.001 &0.001 &0.001 &0.001 &0.001 &0.001 &0.001 &0.001 &0.775 &8.005  & 27.541 \\ 
Mouse&0.001 &0.001 &0.001 &0.001 &0.001 &0.001 &0.001 &0.001 &0.001 &0.059 & 1.195  & 5.834\\ 
Macaque&0.002 &0.002 &0.003 &0.003 &0.003 &0.002 &0.001 &0.002 &0.001 &1.335 & 26.518 & 284.850\\ 
Food&0.004 &0.003 &0.006 &0.005 &0.005 &0.004 &0.002 &0.005 &0.002 &2.672 & 43.386 & 902.276\\ 
Celegans&0.009 &0.006 &0.014 &0.012 &0.012 &0.006 &0.003 &0.013 &0.003 &17.934 & 262.40 & 4117.28\\
USAir&0.007 &0.006 &0.012 &0.011 &0.009 &0.007 &0.002 &0.010 &0.003 &21.763 & 358.76 & 4674.01\\  
Hamster&0.117 &0.112 &0.211 &0.178 &0.162 &0.076 &0.035 &0.215 &0.046 &706.590 &\hspace{0.3cm} - &\hspace{0.4cm} -\\ 
PB&0.131 &0.105 &0.198 &0.190 &0.152 &0.088 &0.041 &0.214 &0.047 &303.530 &\hspace{0.3cm} - &\hspace{0.4cm} -\\ 
Yeast&0.123 &0.116 &0.224 &0.113 &0.166 &0.075 &0.040 &0.094 &0.066 &1133.70 &\hspace{0.3cm} - &\hspace{0.4cm} -\\ 
INT&0.236 &0.150 &0.706 &0.231 &0.507 &0.223 &0.138 &0.146 &0.208 &4956.69 &\hspace{0.3cm} - &\hspace{0.4cm} -\\ 
P2P&0.439 &0.380 &1.280 &0.610 &0.922 &0.412 &0.250 &0.538 &0.387 &7947.25 &\hspace{0.3cm} - &\hspace{0.4cm} -\\ 
Scimet&0.145 &0.143 &0.329 &0.201 &0.251 &0.109 &0.060 &0.191 &0.119 &1862.38 &\hspace{0.3cm} - &\hspace{0.4cm} -\\
\hline
\end{tabular}
\caption{\label{tab:time}Time costs of all compared methods on 12 networks (in seconds). The results are average of 100 independent runs.}
\end{table*}

\section{Conclusions}
\label{sec:Conclusions}
Local clustering information is fundamental and important in many areas, such as in estimating the probability of existing a link in a specific local network. Thus estimating a specific local clustering degree properly becomes a key problem. In this paper, we propose and investigate the power of asymmetric link clustering information in solving the link prediction problem. Previous literatures have verified the effectiveness of node clustering information, but it is still limited. Because traditional node clustering coefficient can not measure the clustering ability of a node to some specific node. To describe the local clustering information more specific, we propose the asymmetric link clustering coefficient. Employing the new measurement, we further give three ALC-based link prediction methods under different frameworks.

To demonstrate the advantage of our methods, we perform globalized and personalized top-L link prediction experiments on 12 networks drawn from various fields. Comparing with NC-based and other non-ALC based methods, ALC-based methods cannot only improve NC-based methods a lot, but also shows great stability in both tasks. While some other good non-ALC based methods, such as LocalPath and CRA indices, can only perform well in one of globalized and personalized link prediction tasks at most. We believe that the main reason, why ALC-based methods achieve such stable improvements, is that asymmetric link clustering measures more targeted local structural information for each candidate pair of nodes. 

At last, we must say there is still a problem we cannot answer, i.e. when ALC-based methods can give definite improvements. Our experiments have shown that ALC-based methods outperform other compared methods in many networks with a significant improvement. However, for USAir network, ALC-based methods fail to perform better than their NC-based counter parts, though they do not perform worse either. Unfortunately, it is not easy to differentiate USAir with other networks from basic network statistics commonly used. Although USAir network shows a significantly high average clustering coefficient, ALC-based methods perform not bad on Macaque network, which has a even higher average clustering coefficient than USAir network. In the future, we will make more efforts to try to answer this question, and we hope that more researchers will adopt ALC-based methods to analyze some other networks. It would be helpful to understand the behavior of asymmetric link clustering information better.

\section{Acknowledgments}

This work is supported by the National Natural Science Foundation of China (Grants No. 61403023), State Education Commission - China Mobile Research Fund (Grants No. MCM20150513) and Fundamental Research Funds for the Central Universities (Grants No. 2017JBM027).

\bibliography{ALCphysicaA}

\end{document}